\documentclass[iop]{emulateapj}

\newcommand\kms{\ifmmode
{\rm km\thinspace s^{-1}}\else km\thinspace s$^{-1}$\fi}
\newcommand\vstar{vB\,120}
\usepackage{lineno}

\shortauthors{Torres et al.}
\shorttitle{\vstar}

\begin{document}
\submitted{Accepted for publication in The Astrophysical Journal}

\title{Dynamical Masses for the Hyades Binary System \vstar}

\author{
Guillermo Torres,
Robert P.\ Stefanik, and
David W.\ Latham
}

\affil{Center for Astrophysics $\vert$ Harvard \& Smithsonian, 60
  Garden St., Cambridge, MA 02138, USA; gtorres@cfa.harvard.edu}

\begin{abstract}
We report spectroscopic observations of \vstar\ (HD~30712), a 5.7~yr
astrometric-spectroscopic binary system in the Hyades cluster. We
combine our radial velocities with others from the literature, and
with existing speckle interferometry measurements, to derive an
improved 3D orbit for the system. We infer component masses of $M_1 =
1.065 \pm 0.018~M_{\sun}$ and $M_2 = 1.008 \pm 0.016~M_{\sun}$, and an
orbital parallax of $21.86 \pm 0.15$~mas, which we show to be more
accurate than the parallax from Gaia~DR3. This is the ninth binary or
multiple system in the Hyades with dynamical mass determinations, and
one of the examples with the highest precision. An analysis of the spectral energy
distribution yields the absolute radii of the stars, $R_1 = 0.968
\pm 0.012~R_{\sun}$ and $R_2 = 0.878 \pm 0.013~R_{\sun}$, and
effective temperatures of $5656 \pm 56$~K and $5489 \pm 60$~
K for the
primary and secondary, respectively. A comparison of these properties
with the predictions of current stellar evolution models for the
known age and metallicity of the cluster shows only minor differences.
\end{abstract}

\section{Introduction}
\label{sec:introduction}

Detached binary systems in which the dynamical masses of the
components can be determined accurately and precisely have long been
used to provide stringent tests of stellar evolution theory.  For
binaries belonging to a cluster of known age and chemical composition,
the constraint is much stronger because one is no longer permitted to
adjust those properties freely (except within their uncertainties) in
order to reach the best agreement between the models and the
observations. The Hyades is among the clusters with the most binary
systems for which masses have been measured (eight to date). One of those systems
\citep[vB\,22 = HD\,27130;][and references therein]{Brogaard:2021} is
eclipsing, and therefore even more useful because it enables the absolute radii
to be determined.\footnote{One other double-lined eclipsing binary in the
Hyades with dynamical mass determinations,
V471\,Tau, is a post-common envelope system consisting of a K-type
main-sequence star and a DA white dwarf \citep[e.g.,][]{Muirhead:2022}. Because
of the prior interaction of the components, it is not suitable for testing
models, as it is not representative of single-star evolution.}
Another relies only on astrometry \citep[vB~80 =
  HD\,28485;][]{Torres:2019}. The others are all
astrometric-spectroscopic systems. These have advantages of their own,
as they provide a model-independent estimate of the distance via the
orbital parallax.

Here we report an orbital analysis for another Hyades binary,
\vstar\ (HD\,30712, $V = 7.73$, \ion{G5}{5}), a recognized member of the cluster
from its proper motion, radial velocity (RV), and parallax. The binary
nature of \vstar\ was announced by \cite{Griffin:1988}, based on
spectroscopic observations started in 1973.  The pair was first
resolved spatially by the technique of speckle interferometry in 1985
\citep{McAlister:1987}, and subsequently also by others.
\cite{Griffin:2012} presented the first double-lined spectroscopic
orbit for the system, featuring a small eccentricity and a period near
5.7~yr. An astrometric orbit that also used the RVs from Griffin was
published by \cite{Tokovinin:2015}, from which the total mass of the
binary was reported as 2.1~$M_{\sun}$. \cite{Docobo:2018} incorporated
a few more speckle measurements, and inferred the individual masses by
adopting the spectroscopic elements of Griffin. The mass uncertainties
were fairly large ($\sim$15\%), however, and were limited mostly by
the astrometric orbit, which suffered from incomplete phase coverage
and a few obvious outliers.

In the interim, several more speckle measurements have become
available, and additionally \vstar\ was monitored at the Center for
Astrophysics (CfA) for 14~yr, as part of a long-running spectroscopic
survey of several hundred stars in the Hyades region. This presents an
opportunity to significantly improve the mass determinations, through
an updated astrometric-spectroscopic orbital analysis. We also aim to infer
the component radii and effective temperatures, from an analysis of the spectral energy distribution aided by the orbital parallax.

The paper is structured as follows. Our RV measurements of \vstar\ are
presented in Section~\ref{sec:spectroscopy}. The astrometric
observations from the literature are described in
Section~\ref{sec:astrometry}, and Section~\ref{sec:orbit} gives the
details of our joint spectroscopic-astrometric orbital analysis. The
fit to the spectral energy distribution of \vstar\ is shown in
Section~\ref{sec:sed}. The photometric variability of the system is
discussed in Section~\ref{sec:activity}, along with other measures of
stellar activity.  The properties we derive for the system (masses,
absolute magnitudes, temperatures, radii) are compared against current
stellar evolution models in Section~\ref{sec:discussion}, and our
conclusions may be found in Section~\ref{sec:conclusions}.

\section{Spectroscopic observations}
\label{sec:spectroscopy}

\vstar\ was observed spectroscopically at the CfA between 1992 October
and 2003 February, with the Digital Speedometer \citep{Latham:1992} on
the 1.5m Wyeth reflector at the (now closed) Oak Ridge Observatory
(Massachusetts, USA). This instrument delivered a resolving power of
$R \approx 35,000$, and was equipped with a photon-counting Reticon
detector that recorded a single echelle order 45~\AA\ wide centered
near 5187~\AA. The main spectral feature in this region is the \ion{Mg}{1}\,b
triplet. Signal-to-noise ratios for the 40 usable exposures range
between 11 and 29 per resolution element of 8.5~\kms. Wavelength
solutions relied on exposures of a thorium-argon lamp taken before and after
each science exposure. The zeropoint of the instrument was monitored
by means of exposures of the twilight sky in the evening and morning.
Those observations
were used to calculate and apply small (typically $\leq 2~\kms$)
run-to-run corrections to the raw velocities we describe next, in order
to place them on a uniform system \citep[see][]{Latham:1992}. This
native CfA system is slightly offset from the IAU reference frame by
0.14~\kms\ \citep{Stefanik:1999}, as determined from observations of
minor planets in the solar system. In order to remove this shift, we
adjusted the velocities by adding +0.14~\kms.

Radial velocities for both components were measured with the
two-dimensional cross-correlation algorithm TODCOR
\citep{Zucker:1994}. We used synthetic templates from a large library
of pre-computed spectra based on model atmospheres by R.\ L.\ Kurucz,
and a line list manually adjusted to provide a better match to real
stars \citep[see][]{Nordstrom:1994, Latham:2002}. These models
adopt a microturbulent velocity of $\xi = 2~\kms$, along with a
macroturbulent velocity of $\zeta_{\rm RT} = 1~\kms$. Grids of
correlations for a range of template parameters gave the highest
average correlation for effective temperatures of 5500~K and a total
line broadening of 6~\kms\ for both stars. The line broadening is
dominated by rotation ($v \sin i$), but includes any difference in the
true macroturbulent velocity compared to the value built into the
templates.  Surface gravities were held at
$\log g = 4.5$ for both stars, and the metallicity was solar, which is
close to the composition of the Hyades \citep[${\rm [Fe/H]} =
  +0.18$;][{see also Section~\ref{sec:discussion}}]{Dutra-Ferreira:2016}.

In our experience, the narrow spectral window of the observations can
occasionally produce systematic errors in the velocities, which are
caused by lines of the components shifting in and out of the spectral
order in opposite directions as a function of orbital phase. We
corrected these effects through numerical simulations
\citep[see][]{Latham:1996, Torres:1997}. The corrections were smaller
than 0.2~\kms, on average, but were applied nonetheless.
Table~\ref{tab:rvs} lists the final velocities and their formal
uncertainties.

\setlength{\tabcolsep}{6pt}
\begin{deluxetable}{lccc}
\tablewidth{0pc}
\tablecaption{CfA Radial Velocity Measurements for \vstar \label{tab:rvs}}
\tablehead{
\colhead{HJD} &
\colhead{$RV_1$} &
\colhead{$RV_2$} &
\colhead{Phase}
\\
\colhead{(2,400,000+)} &
\colhead{(\kms)} &
\colhead{(\kms)} &
\colhead{}
}
\startdata
  48904.8585  &  $37.33 \pm 1.11$  &  $46.92 \pm 1.20$  & 0.0797 \\
  48936.8503  &  $36.32 \pm 0.59$  &  $48.82 \pm 0.64$  & 0.0949 \\
  48990.5801  &  $34.19 \pm 1.05$  &  $49.26 \pm 1.13$  & 0.1205 \\
  49048.6166  &  $33.44 \pm 0.88$  &  $50.62 \pm 0.95$  & 0.1481 \\
  49260.8359  &  $32.74 \pm 0.68$  &  $52.52 \pm 0.73$  & 0.2492 
\enddata
\tablecomments{Orbital phases in the last column are computed from
the ephemeris in Table~\ref{tab:mcmc}. (This table is available in
its entirety in machine-readable form)}
\end{deluxetable}
\setlength{\tabcolsep}{6pt}

More precise temperatures for the \vstar\ components were obtained by
interpolation among the templates, following \cite{Torres:2002}.  This
resulted in values of 5620 and 5450~K for the primary and secondary,
with estimated uncertainties of 100~K. As the true metallicity of the
cluster is slightly supersolar, we repeated this exercise for ${\rm
  [Fe/H]} = +0.5$, which gave temperatures about 300~K hotter for each
star. Interpolation to ${\rm [Fe/H]} = +0.18$ then led to final values
of 5730 and 5560~K ($\pm 100$~K).

Using TODCOR, we obtained a flux ratio between the components of
$\ell_2/\ell_1 = 0.664 \pm 0.026$. This corresponds to a magnitude
difference of $\Delta m = 0.44 \pm 0.04$ at the wavelength of our
observations. We use this estimate later in Section~\ref{sec:sed}.

In addition to our own RVs, the orbital analysis described below
incorporated the 41 pairs of velocities of \cite{Griffin:2012}, which are of
similar precision as ours. We adopted relative weights for those
measurements as recommended by the author, along with the specified
error for an observation of unit weight. Separate spectroscopic
orbital solutions using the Griffin observations and our own give
similar velocity semiamplitudes.

\section{Astrometric measurements}
\label{sec:astrometry}

Speckle observations of \vstar\ have been recorded by several authors
since it was first resolved in 1985. A listing of all measurements
from the Washington Double Star Catalog \citep[WDS;][]{Worley:1997,
  Mason:2001} was kindly provided by R.\ Matson (U.S.\ Naval
Observatory). The double star designation in this catalog is
WDS~J04506+1505AB.  While many of the early measurements suffer from a
180\arcdeg\ ambiguity in the position angles, the most recent (2021)
observations by \cite{Tokovinin:2022} were reduced with a methodology
that is able to distinguish the correct quadrant. All other position
angles were then changed as needed. After consulting the original
sources, minor adjustments were made also to some of the uncertainties in order to
make them more realistic, as they often account only for the internal
errors.  Observations published with no indication of their precision
were assigned initial errors of 1\arcdeg\ in position angle ($\theta$)
and 2~mas in the separation ($\rho$). All uncertainties were later
adjusted during the analysis, as we describe below. Three WDS
measurements from 1991.9023, 2014.8568, and 2005.8688 were found to
give abnormally large residuals, and were excluded.
Table~\ref{tab:speckle} gives the final list of speckle observations
as used here.

\setlength{\tabcolsep}{5pt}
\begin{deluxetable}{lcccc}
\tablewidth{0pc}
\tablecaption{Speckle Measurements for \vstar \label{tab:speckle}}
\tablehead{
\colhead{Year} &
\colhead{$\theta$} &
\colhead{$\rho$} &
\colhead{Phase} &
\colhead{Source}
\\
\colhead{} &
\colhead{(degree)} &
\colhead{(\arcsec)} &
\colhead{} &
\colhead{}
}
\startdata
  1985.8459  &  \phn$290.2 \pm 1.0$\phn\phn  &   $0.072 \pm 0.002$   &  0.8748  &   1   \\ 
  1988.2601  &  \phn$120.5 \pm 1.0$*\phn     &   $0.085 \pm 0.002$   &  0.2948  &   2   \\ 
  1988.6609  &  \phn$107.1 \pm 1.0$*\phn     &   $0.071 \pm 0.002$   &  0.3645  &   3   \\ 
  1990.7554  &  \phn$314.7 \pm 1.0$\phn\phn  &   $0.090 \pm 0.002$   &  0.7289  &   4   \\ 
  1993.8420  &  \phn$127.8 \pm 2.0$\phn\phn  &   $0.090 \pm 0.005$   &  0.2658  &   5   \\ 
  1996.025   &    \phn$329 \pm 1$*\phn       &   $0.069 \pm 0.002$   &  0.6456  &   6   \\ 
  1996.8689  &  \phn$301.4 \pm 2.0$\phn\phn  &   $0.082 \pm 0.004$   &  0.7924  &   7   \\ 
  2010.9658  &  \phn$126.3 \pm 0.5$\phn\phn  &  $0.0902 \pm 0.0005$  &  0.2448  &   8   \\ 
  2010.9658  &  \phn$127.3 \pm 0.6$\phn\phn  &  $0.0898 \pm 0.0005$  &  0.2448  &   8   \\ 
  2015.7438  &  \phn$163.3 \pm 0.9$*\phn     &  $0.0544 \pm 0.0009$  &  0.0760  &   9   \\ 
  2015.9104  &  \phn$154.0 \pm 0.5$*\phn     &  $0.0651 \pm 0.0009$  &  0.1050  &   9   \\ 
  2016.9573  &  \phn$120.9 \pm 0.7$*\phn     &  $0.0865 \pm 0.0006$  &  0.2871  &   10  \\ 
  2017.9320  &   \phn$69.7 \pm 0.5$*         &  $0.0414 \pm 0.0005$  &  0.4566  &   11  \\ 
  2018.8409  &  \phn$336.0 \pm 0.5$*\phn     &  $0.0644 \pm 0.0005$  &  0.6148  &   11  \\ 
  2019.7914  &  \phn$307.0 \pm 0.5$*\phn     &  $0.0882 \pm 0.0005$  &  0.7801  &   12  \\ 
  2020.8346  &  \phn$252.9 \pm 0.6$*\phn     &  $0.0374 \pm 0.0005$  &  0.9616  &   13  \\ 
  2021.7982  &  \phn$146.7 \pm 0.5$\phn\phn  &  $0.0708 \pm 0.0005$  &  0.1292  &   14  \\ 
  2021.7982  &  \phn$146.6 \pm 0.5$\phn\phn  &  $0.0702 \pm 0.0005$  &  0.1292  &   14  
\enddata
\tablecomments{Position angles $\theta$ are given for the equinox of
  the date of the observation. Asterisks indicate angles we have
  changed by 180\arcdeg\ to place them in the proper quadrant. Orbital
  phases are based on the ephemeris of
  Table~\ref{tab:mcmc}. Sources in the last column are:
(1) \cite{McAlister:1987};
(2) \cite{McAlister:1989};
(3) \cite{McAlister:1990};
(4) \cite{Hartkopf:1992};
(5) \cite{Balega:1994};
(6) \cite{Patience:1998};
(7) \cite{Hartkopf:2000};
(8) \cite{Hartkopf:2012};
(9) \cite{Tokovinin:2016};
(10) \cite{Tokovinin:2018};
(11) \cite{Tokovinin:2019};
(12) \cite{Tokovinin:2020};
(13) \cite{Tokovinin:2021};
(14) \cite{Tokovinin:2022}.
}
\end{deluxetable}
\setlength{\tabcolsep}{6pt}

\vskip 20pt
\section{Orbital analysis}
\label{sec:orbit}

A joint analysis of the CfA and Griffin velocities, and of the speckle
measurements, was performed within a Markov chain Monte Carlo (MCMC)
framework using the {\sc emcee}
package\footnote{\url{https://emcee.readthedocs.io/en/stable/index.html}}
of \cite{Foreman-Mackey:2013}. All position angles were adjusted for
precession to the year J2000. We solved for the orbital period ($P$),
the angular semimajor axis ($a^{\prime\prime}$), the cosine of the
inclination angle ($\cos i$), the position angle of the ascending node
for J2000.0 ($\Omega$), the eccentricity ($e$) and argument of
periastron for the primary ($\omega_1$), cast as
$\sqrt{e}\cos\omega_1$ and $\sqrt{e}\sin\omega_1$, a reference time of
periastron passage ($T_{\rm peri}$), and the spectroscopic parameters
$K_1$, $K_2$, and $\gamma$, which are the velocity semiamplitudes and
center-of-mass velocity. An additional free parameter $\Delta_{\rm G}$
was included to allow for a possible systematic offset between the CfA
and Griffin velocities. It corresponds to the correction to be added
to the Griffin RVs in order to place them on the CfA scale.  To ensure
proper weighting of the observations, six additional parameters were
included in the analysis to represent multiplicative scaling factors
for the formal uncertainties, which are not always accurate: $f_{\theta}$ and
$f_{\rho}$ for the astrometric measurements, $f_{\rm C,1}$ and $f_{\rm
  C,2}$ for the primary and secondary velocities from CfA, and $f_{\rm
  G,1}$ and $f_{\rm G,2}$ for the Griffin RVs. The total number of
free parameters was 17.

We used 100 random walkers, and the MCMC chains had 10,000 links each, after
burn-in. Convergence was checked by visual examination of the chains,
and by requiring a Gelman-Rubin statistic of 1.05 or smaller
\citep{Gelman:1992}. Priors were all uniform over suitable ranges,
except for those of the error scaling factors, which were log-uniform.

\setlength{\tabcolsep}{5pt}
\begin{deluxetable}{lcc}
\tablewidth{0pc}
\tablecaption{Orbital Parameters for \vstar \label{tab:mcmc}}
\tablehead{
\colhead{Parameter} &
\colhead{Value} &
\colhead{Prior}
}
\startdata
 $P$ (day)                         & $2099.57 \pm 0.60$\phm{222}  & [1800, 2300]   \\
 $T_{\rm peri}$ (HJD$-$2,400,000)  & $52937 \pm 10$\phm{222}      & [52300, 53500] \\
 $a^{\prime\prime}$ (arcsec)       & $0.08943 \pm 0.00035$        & [0.02, 0.20]   \\
 $\sqrt{e}\cos\omega_1$            & $-0.0046 \pm 0.0072$\phs     & [$-1$, 1]        \\
 $\sqrt{e}\sin\omega_1$            & $+0.2350 \pm 0.0068$\phs     & [$-1$, 1]        \\
 $\cos i$                          & $-0.3948 \pm 0.0047$\phs     & [$-1$, 1]        \\
 $\Omega$ (degree)                 & $129.59 \pm 0.30$\phn\phn    & [0, 360]       \\
 $K_1$ (\kms)                      & $9.484 \pm 0.062$            & [2, 15]        \\
 $K_2$ (\kms)                      & $10.026 \pm 0.072$\phn       & [2, 15]        \\
 $\gamma$ (\kms)                   & $+42.112 \pm 0.057$\phn\phs  & [30, 60]       \\
 $\Delta_{\rm G}$ (\kms)           & $-0.824 \pm 0.074$\phs       & [$-5$, 5]        \\
 $f_{\theta}$                      & $1.77 \pm 0.45$              & [$-5$, 5]        \\
 $f_{\rho}$                        & $2.25 \pm 0.53$              & [$-5$, 5]        \\
 $f_{\rm C,1}$                     & $0.660 \pm 0.087$            & [$-5$, 5]        \\
 $f_{\rm C,2}$                     & $0.702 \pm 0.092$            & [$-5$, 5]        \\
 $f_{\rm G,1}$                     & $1.05 \pm 0.14$              & [$-5$, 5]        \\
 $f_{\rm G,2}$                     & $1.09 \pm 0.15$              & [$-5$, 5]        \\ [1ex]
\hline \\ [-1.5ex]
\multicolumn{3}{c}{Derived Properties} \\ [0.5ex]
\hline \\ [-1.5ex]
 $i$ (degree)                      & $113.25 \pm 0.29$\phn\phn    & \nodata  \\
 $e$                               & $0.0552 \pm 0.0033$          & \nodata  \\
 $\omega_1$ (degree)               & $91.1 \pm 1.8$\phn           & \nodata  \\
 $a$ (au)                          & $4.091 \pm 0.022$            & \nodata  \\
 $M_1$ ($M_{\sun}$)                & $1.065 \pm 0.018$            & \nodata  \\
 $M_2$ ($M_{\sun}$)                & $1.008 \pm 0.016$            & \nodata  \\
 $q \equiv M_2/M_1$                & $0.9460 \pm 0.0094$          & \nodata  \\
 $\pi_{\rm orb}$ (mas)             & $21.86 \pm 0.15$\phn         & \nodata  \\
 Distance (pc)                     & $45.75 \pm 0.32$\phn         & \nodata 
\enddata

\tablecomments{The values listed correspond to the mode of the
  posterior distributions, with uncertainties representing the 68.3\%
  credible intervals. Priors in square brackets are uniform over the
  ranges specified, except those for the error inflation factors $f$,
  which are log-uniform.}

\end{deluxetable}
\setlength{\tabcolsep}{6pt}

\begin{figure}
\epsscale{1.17}
\plotone{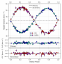}

\figcaption{Radial-velocity measurements for \vstar\ from the CfA and
  \cite{Griffin:2012}, along with our adopted model for the
  spectroscopic orbit. The dotted line at the top represents the
  center-of-mass velocity. Residuals are shown at the bottom. Phase
  0.0 corresponds to periastron passage.\label{fig:rvs}}

\end{figure}

\begin{figure}
\epsscale{1.17}
\plotone{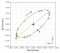}

\figcaption{Speckle observations for \vstar\ with our adopted model
  for the astrometric orbit. The ``+" sign represents the primary
  star. Short line segments connect the measured
  position with the predicted location of the secondary on the orbit. The line of nodes
  is indicated with a dotted line, and $\Omega$ marks the ascending
  node. The location of periastron is indicated with the square marked
  ``P''. \label{fig:visual}}

\end{figure}

The results of the analysis are presented in Table~\ref{tab:mcmc}.
Derived properties are given at the bottom, and include the masses and
the orbital parallax, $\pi_{\rm orb}$. A graphical representation of the spectroscopic
orbit is shown in Figure~\ref{fig:rvs}, and the speckle orbit can be
seen in Figure~\ref{fig:visual}.

In addition to the present orbital solution and that of
\cite{Griffin:2012}, previous ones for \vstar\ include a preliminary
astrometric analysis of the Hipparcos data by \cite{Soderhjelm:1999},
which assumed a circular orbit, an astrometric-spectroscopic analysis
somewhat similar to ours by \cite{Tokovinin:2015}, in which they
reported only the astrometric elements, and an astrometric solution by
\cite{Docobo:2018}, in which several of the elements were held fixed
from the work of \cite{Griffin:2012}.  Table~\ref{tab:orbits} presents
a comparison of all of these results.

\setlength{\tabcolsep}{6pt}
\begin{deluxetable*}{lccccc}
\tablewidth{0pc}
\tablecaption{Orbital Parameters for \vstar\ from this Work Compared with Previous Determinations \label{tab:orbits}}
\tablehead{
\colhead{Parameter} &
\colhead{\cite{Soderhjelm:1999}} &
\colhead{\cite{Griffin:2012}} &
\colhead{\cite{Tokovinin:2015}} &
\colhead{\cite{Docobo:2018}} &
\colhead{This work}
}
\startdata
$P$ (year)                   &     7.5      &  $5.7342 \pm 0.0057$        &     $5.734 \pm 0.005$          &    fixed                 &    $5.7483 \pm 0.0016$         \\
$a^{\prime\prime}$ (arcsec)  &    0.096     &     \nodata                 &    $0.0904 \pm 0.0010$         & $0.089 \pm 0.003$        &   $0.08943 \pm 0.00035$        \\
$e$                          &   0 (fixed)  &   $0.066 \pm 0.006$         &     $0.058 \pm 0.007$          &    fixed                 &    $0.0552 \pm 0.0033$         \\
$i$ (degree)                 &       71     &     \nodata                 &     $114.4 \pm 1.7$\phn\phn    & $116.7 \pm 1.5$\phn\phn  &    $113.25 \pm 0.29$\phn\phn   \\
$\omega_1$ (degree)          &    \nodata   &      $96 \pm 6$\phn         &      $92.8 \pm 6.0$\phn        &    fixed                 &      $91.1 \pm 1.8$\phn        \\
$\Omega$ (degree)            &      132     &     \nodata                 &     $130.1 \pm 0.7$\phn\phn    & $130.5 \pm 1.0$\phn\phn  &    $129.59 \pm 0.30$\phn\phn   \\
$T_{\rm peri}$ (yr)          &  1994.3      & $2003.81 \pm 0.10$\phm{222} &  $2003.829 \pm 0.096$\phm{222} &    fixed                 &  $2003.810 \pm 0.027$\phm{222} \\
$K_1$ (\kms)                 &    \nodata   &    $9.54 \pm 0.08$          &     \nodata                    &  \nodata                 &     $9.484 \pm 0.062$          \\
$K_2$ (\kms)                 &    \nodata   &    $9.89 \pm 0.08$          &     \nodata                    &  \nodata                 &    $10.026 \pm 0.072$\phn      \\
$\gamma$ (\kms)              &    \nodata   &   $42.96 \pm 0.04$\phn      &     \nodata                    &  \nodata                 &    $42.112 \pm 0.057$\phn      
\enddata
\tablecomments{The elements held fixed in the Docobo solution were taken from the
work of \cite{Griffin:2012}.}
\end{deluxetable*}

\section{The spectral energy distribution (SED)}
\label{sec:sed}

\vstar\ has been observed in a variety of standard photometric
systems. These measurements of the combined light can be used to
derive estimates of the effective temperatures and angular diameters
of the components, by comparison with appropriate synthetic spectra
based on model atmospheres. Scaling the angular diameters by the distance
derived from the
orbital parallax obtained in the previous section then provides the
absolute radii. We collected a total of 44 individual brightness
measurements from the VizieR
database\footnote{\url{https://vizier.cds.unistra.fr/viz-bin/VizieR}},
in the following photometric systems:
Johnson ($U, B, V, R, I$), Tycho-2 ($B_{\rm T}, V_{\rm T}$), Hipparcos
($Hp$), Gaia DR3 ($G, G_{\rm BP}, G_{\rm RP}$), Geneva ($U, B1, B, B2, V1, V,
G$), Pan-STARRS ($g, r, z$), the WBVR system ($W, B, V, R$), Str\"omgren
($u, v, b, y$), WISE ($W1$--$W4$), GALEX ($NUV$), and 2MASS ($J, H, K_{\rm
  S}$).
Together, these observations span the entire optical range, and extend
also into the UV and infrared (0.25--22~$\mu$m). To further constrain
the SED fit, we made use of the available estimates of the magnitude
difference between the stars at different wavelengths. From our own
spectroscopic observations in Section~\ref{sec:spectroscopy}, we
inferred $\Delta m = 0.44 \pm 0.04$~mag in the 5187~\AA\ region
covered by our spectra. A further estimate of $\Delta m \approx
0.37$~mag was reported by \cite{Griffin:2012}, approximately in the
$V$ band. We assign it an uncertainty of 0.02~mag. Other measurements
have been obtained by the speckle observers. The weighted average of
those estimates in $V$ is $0.37 \pm 0.11$~mag, and a measurement in
the $K$ band gave $\Delta m = 0.25 \pm 0.03$~mag.

As the SED is insensitive to the stars' surface gravities,
we adopted typical values for dwarfs of
$\log g = 4.5$. The metallicity was held fixed at a
value appropriate for the Hyades \citep[$\rm{[Fe/H]} =
  +0.18$;][{see below}]{Dutra-Ferreira:2016}. In addition to the individual
temperatures and angular diameters, we solved for an additional
photometric error added in quadrature to the published uncertainties of all
the measured magnitudes, to account for possible underestimates of the
observational errors as well as uncertainties in the various
photometric zeropoints and calibrations. We also allowed the reddening
$E(B-V)$ to be free, even though it is typically considered to be
negligible for the Hyades \citep[see, e.g.,][]{Taylor:2006}.

\begin{figure}
\epsscale{1.17}
\plotone{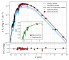}

\figcaption{Spectral energy distribution for \vstar, fitted using
  Kurucz stellar atmosphere models \citep{Castelli:2003}. The
  horizontal error bars represent the width of each bandpass. Residuals
  at the bottom are shown in units of magnitudes. The additional
  photometric error required by our analysis to achieve a reduced
  $\chi^2$ value of unity is $0.025 \pm 0.004$~mag.  The inset shows
  the predicted magnitude difference between the components from the
  best-fit model, along with the available $\Delta m$
  measurements described in the text. \label{fig:sed}}

\end{figure}

We illustrate our fit in Figure~\ref{fig:sed}. We obtained
temperatures of $5656 \pm 56$~K and $5489 \pm 60$~K for the primary
and secondary, where we have conservatively added 50~K in quadrature
to the internal uncertainties to allow for the possibility of
systematic errors in the models. The resulting angular diameters are
$0.1969 \pm 0.0020$~mas and $0.1786 \pm 0.0025$~mas, leading to
corresponding absolute radii of $0.968 \pm 0.012~R_{\sun}$ and $0.878
\pm 0.013~R_{\sun}$. The radius uncertainties include the contribution
from the formal error in the orbital parallax. The temperatures are
slightly lower than our spectroscopic values by about 70~K, but are
formally more precise. As expected, the reddening we infer is
consistent with zero: $E(B-V) = 0.0047 \pm 0.0053$~mag.

\section{Stellar activity}
\label{sec:activity}

\vstar\ has long been listed as a suspected variable star in the
literature, with the designation NSV~1735. However, as pointed out by
\cite{Griffin:2012}, this seems to have been a consequence of a
possible misprint in a single measurement of the $V$ magnitude in the
1950s ($V = 7.34$), which had it about 0.4~mag brighter than all
other determinations. No other reliable measurement as bright as
this has been published since.

As it turns out, the very high photometric precision, high cadence,
and continuity now attainable with space-based missions such as
Kepler/K2 and TESS have shown that most stars are variable at some
level, and \vstar\ is no exception. \cite{Douglas:2019} used data from
the K2 mission, and inferred a photometric period of 8.61~d,
presumably due to rotation. \cite{Green:2023} examined the TESS
observations, and reported a period of 4.41~d, about half as long.

\begin{figure}
\epsscale{1.17}
\plotone{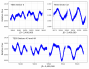}

\figcaption{TESS light curve of \vstar\ in the four sectors observed
  by the satellite as of this writing.\footnote{Data downloaded from
the Mikulski Archive for Space Telescopes, \url{https://archive.stsci.edu/}}.
    Fluxes in each sector have been normalized by the median value. \label{fig:tess}}

\end{figure}

Figure~\ref{fig:tess} shows the measurements from TESS for the four
sky sectors currently available (5, 32, 43, and 44), in which the
peak-to-peak variation over this interval is only about 1.5~mmag.
\vstar\ is the brightest object in the photometric aperture.
Sector~32 clearly shows a single dominant period of about 8.5~d,
consistent with the estimate of \cite{Douglas:2019}. Sector~5, the one
on which \cite{Green:2023} based their result, as well as sectors 43
and 44, display a more complicated structure. Given the similarity of
the properties of the two components, including their brightness, it
seems plausible that both stars contribute to the variability
with similar periods of roughly 8.5~d. In that case, it is possible that the corresponding
peaks and troughs in the lightcurve happened to line up for the two
stars in sector~32 (or the spots on one star disappeared),
but were out of phase in the other sectors. We
speculate that this misalignment could have led to the shorter period
reported by \cite{Green:2023} from sector~5. Alternatively, perhaps
only one star is spotted, and its surface features come and go,
or has two spots on opposite sides, one of which comes and goes.

If the rotation periods are both indeed $\sim$8.5~d, the projected
rotational velocities would be expected to be $\sim$5.5\,$\sin i_{\rm rot}~\kms$ for an
average radius of 0.9~$R_{\sun}$, or about 5~\kms\ if the spin axes
are parallel to the orbital axis ($i_{\rm rot} \approx i =
113\fdg25$). This value is not far from the adopted 6~\kms\ line
broadening in our spectroscopic analysis of
Section~\ref{sec:spectroscopy}. It is also close to the $v \sin i$
estimate of \cite{Griffin:2012} for the primary ($4.6 \pm 0.7~\kms$),
although his estimate for the secondary is much lower ($2.1 \pm
0.9~\kms$). A shorter rotation period of 4.41~d \citep{Green:2023} would require stronger
spin/orbit misalignments to match the measured line
broadening.

\vstar\ is a known X-ray source, having been detected by the ROSAT and
XMM-Newton missions.  The strength of its chromospheric activity as
measured by the emission cores in the \ion{Ca}{2} H and K lines has
been reported as $\log R^\prime_{\rm HK} = -4.52$ \citep[][average of
  three measurements]{Isaacson:2010} or $\log R^\prime_{\rm HK} =
-4.35$ \citep{Brown:2022}. While higher than the Sun, this level of
activity is typical of Hyades stars with the spectral type of \vstar.
Based on a similar $\log R^\prime_{\rm HK}$ measure,
\cite{Fuhrmeister:2022} predicted a rotation period of 9.2~d using a
statistical relation by \cite{Mittag:2018}, which is rather close to
the period seen in the TESS data.

\section{Discussion}
\label{sec:discussion}

{The absolute masses of \vstar, and its other physical properties,
offer an opportunity for a comparison against current stellar evolution
models at the age and composition of the Hyades. Both of these cluster attributes
have some degree of uncertainty, however. For the age, a commonly quoted
value is that of \cite{Perryman:1998}, $625 \pm 50$~Myr, obtained by
isochrone modeling of the color-magnitude diagram (CMD) using the
Hipparcos parallaxes. A very similar
estimate of $650 \pm 70$~Myr, based on the lithium depletion boundary, was
reported by \cite{Martin:2018}. It has been found that rotation 
can alter the shape of the upper main-sequence in the CMD, thereby
affecting cluster ages.
Accounting for this, \cite{Brandt:2015a} derived a rather older 
age for the Hyades of about 800~Myr, subsequently revised to
$750 \pm 100$~Myr \citep{Brandt:2015b}.
A similar analysis by \cite{Gossage:2018} yielded an age of 680~Myr from
the optical CMD, which they favored over a larger estimate of about 740~Myr in
the near infrared. For a compilation of other age estimates, see
\cite{Douglas:2019}.

Metallicity determinations for the Hyades have typically ranged between
${\rm [Fe/H]} = +0.1$ and +0.2~dex. Classical spectroscopic studies of
FGK-type dwarfs find a metallicity of about +0.13 or +0.14~dex
\citep[e.g.,][]{Cayrel:1985, Boesgaard:1990, Paulson:2003,
Schuler:2006}. Estimates for the giants give similar values \citep{Smith:1999,
Carrera:2011, Ramya:2019}. Variations from star to star have been
reported, particularly among the A stars, as well as systematic differences
between \ion{Fe}{1} and \ion{Fe}{2} abundances, increasing toward the
cooler stars \citep[e.g.,][]{Aleo:2017}.
\cite{Dutra-Ferreira:2016} examined the importance of the
line list, the consistency between dwarfs and giants, and different
ways of constraining other stellar parameters needed for the analysis
($T_{\rm eff}$, $\log g$, microturbulent velocity). They expressed a
preference for the value ${\rm [Fe/H]} = +0.18 \pm 0.03$, which they
found to be the same for dwarfs and giants, and to be robust against
the choice of the line list.

For the purposes of this paper, we adopt a Hyades metallicity of
${\rm [Fe/H]} = +0.18$ from \cite{Dutra-Ferreira:2016}, and an age
of 750~Myr from \cite{Brandt:2015b}. The impact of these choices
is discussed below.
}

\begin{figure}
\epsscale{1.17}
\plotone{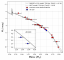}
\figcaption{Mass-luminosity relation for the Hyades cluster in the
  visual band. The masses for \vstar\ are from Table~\ref{tab:mcmc},
  and the absolute magnitudes are based on a typical system brightness
  of $V = 7.73 \pm 0.02$, a $V$-band magnitude difference of $0.37 \pm
  0.02$ \citep{Griffin:2012}, the orbital parallax from the present
  work, and the assumption of zero extinction. Measurements for the
  other systems are taken from \cite{Torres:2019}, with updates for
  two of them by \cite{Brogaard:2021} and \cite{Anguita-Aguero:2022},
  and from \cite{Torres:2023}.
  For comparison, model isochrones from two different series of
  calculations are also shown, for the age
  \cite[750~Myr;][]{Brandt:2015b} and metallicity of the cluster. The
  inset shows a close-up of \vstar.\label{fig:VbandM-L}}

\end{figure}

\begin{figure}
\epsscale{1.17}
\plotone{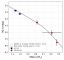}
\figcaption{Similar to Figure~\ref{fig:VbandM-L}, for the $K$ band. The
  individual absolute magnitudes for \vstar\ use the system brightness
  from 2MASS, and the magnitude difference of $0.25 \pm 0.03$ from
  \cite{Patience:1998}. HD~284163 (also shown) is the only other
  multiple system in the Hyades with known masses and individual $K$-band
  magnitudes \citep{Torres:2023}. \label{fig:KbandM-L}}

\end{figure}

Figure~\ref{fig:VbandM-L} shows the empirical mass-luminosity relation
for the Hyades in the visual band. It includes the components of all
binary systems with previously measured dynamical masses. \vstar\ is
in general agreement with the trend (slope), although both components
fall slightly below the stellar evolution models shown in the figure
for the {adopted} age and metallicity of the Hyades, by about 1 or 1.5$\sigma$
(see the inset). Other stars between 1 and 1.5~$M_{\sun}$ also tend to
be slightly fainter than predicted by theory, on average. One
possible explanation, though it seems unlikely, might be systematic errors affecting the
magnitudes and/or masses of several of these systems in the same way.
Another could be missing opacities in the stellar atmosphere models
used to compute the fluxes for the isochrones. A similar diagram for
the $K$ band (Figure~\ref{fig:KbandM-L}) indicates that the primary of
\vstar\ is consistent with both the MIST model isochrone of
\cite{Choi:2016} and the PARSEC~v1.2S model of \cite{Chen:2014}, while
the secondary is marginally below the latter. The one other system in
the Hyades for which the individual masses and $K$-band absolute
magnitudes are known is HD~284163 \citep{Torres:2023}, a system of
three stars also shown in the figure.

\begin{figure}
\epsscale{1.17}
\plotone{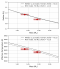}

\figcaption{Individual radii and effective temperatures for the
components of \vstar\ from our
SED fit, compared with the same stellar evolution models as in
Figures~\ref{fig:VbandM-L} and \ref{fig:KbandM-L}.
\label{fig:radteff}}

\end{figure}

A comparison between those same models and the components'
effective temperatures and radii, from our SED analysis of
Section~\ref{sec:sed}, is
shown in Figure~\ref{fig:radteff}. The radii are consistent with the
models, within their uncertainties. The effective temperatures agree
well with the MIST isochrone, and are just slightly cooler but
essentially also within 1$\sigma$ of the PARSEC~v1.2S models, which are
in turn systematically hotter than MIST by about 70~K. 

\begin{figure}
\epsscale{1.17}
\plotone{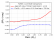}

\figcaption{{Influence of the age and metallicity on the absolute
$V$-band magnitude predicted by the PARSEC~v1.2S models for the
Hyades, as a function of mass. 
The dashed line shows changes in $M_V$ for different compositions, at a fixed age of
750~Myr (our reference age). The solid line shows changes for different ages at a fixed
reference composition of ${\rm [Fe/H]} = +0.18$. Differences between
models are taken in the sense indicated in the labels.} \label{fig:comparemodels}}

\end{figure}

{Adopting a different age and/or composition for the Hyades
changes the predicted absolute magnitudes from the models by a
small but non-negligible amount that is mass-dependent.
Figure~\ref{fig:comparemodels}
quantifies this for the PARSEC~v1.2S models in the $V$ band,
addressing the comparison shown previously in Figure~\ref{fig:VbandM-L}.
For example, lowering the metallicity from our chosen value of ${\rm [Fe/H]} = +0.18$
to +0.13~dex at a fixed age
(dashed line in
Figure~\ref{fig:comparemodels}) makes the models slightly brighter
(negative $\Delta M_V$). The effect is roughly 0.04~mag for masses
larger than about 1.2~$M_{\sun}$, gradually increasing toward lower masses.
On the other hand, reducing the age from 750~Myr to 625~Myr at a fixed metallicity
makes the predicted magnitudes fainter (positive
$\Delta M_V$), by up to 0.04~mag for unevolved stars with masses
below about 1.7~$M_{\sun}$. For higher masses the differences are larger,
as such stars begin to evolve (solid line). 

Overall, we estimate that the uncertainty in the age and composition of the
Hyades can affect the comparison in Figure~\ref{fig:VbandM-L} at
a level similar to the observational errors in $M_V$.
}

The orbital parallax of \vstar\ from our analysis, $\pi_{\rm orb} = 21.86 \pm
0.15$~mas, has a formal error of only 0.7\%. The entry for the object in
the Gaia~DR3 catalog \citep[source identifier
  3404812685132622592;][]{GaiaDR3:2022} is lower ($20.53 \pm
0.18$~mas)\footnote{We include here a zeropoint correction of
  +0.03~mas, as advocated by \cite{Lindegren:2021}.}, but did not
account for the orbital motion. A sign of this is evident in the reported
quality of the astrometric fit, as measured by the Renormalized Unit
Weight Error (RUWE). For well-behaved sources in which a single-star
model provides a good fit, the RUWE is expected to be near 1.0. Values
larger than about 1.4 could be indicative of a non-single source, or
an otherwise problematic astrometric solution. \vstar\ has a RUWE of
7.116. While this does not necessarily imply the parallax is biased,
at the very least the formal uncertainty will be underestimated.
Following the prescription by \cite{MaizApellaniz:2022}, we estimate
the external error of the Gaia~DR3 parallax to be 0.63~mas. Even with
this larger error, the difference compared to our orbital parallax is
still at the 2$\sigma$ level, likely a consequence of unmodeled
binary motion in Gaia.

The Hipparcos mission delivered a parallax for \vstar\ of $23.64 \pm
0.99$~mas \citep[source identifier HIP~22505;][]{vanLeeuwen:2007},
which is about 2$\sigma$ larger than ours. This also did not account
for orbital motion.  \cite{Madsen:2002} applied the moving-cluster
method to the Hyades using Hipparcos positions and proper motions, and
obtained purely kinematic parallaxes for many of the cluster
members that are often better than the trigonometric values.
Their result for \vstar\ is $21.85 \pm 0.39$~mas, which is
rather different from the trigonometric value from the mission, but is
nearly identical to ours.  The study also derived an astrometric
radial velocity for \vstar\ of $41.81 \pm 0.60~\kms$, again very close
to the $\gamma$ velocity listed in our Table~\ref{tab:mcmc}. A similar
study using astrometry from the Gaia~DR1 catalog \citep{GaiaDR1:2016}
was carried out by \cite{Reino:2018}, who reported a kinematic
parallax of $21.58 \pm 0.23$~mas, along with an astrometric RV of
$41.97 \pm 1.05~\kms$.  Both are consistent with the results of this
paper.

\section{Conclusions}
\label{sec:conclusions}

In this work we have made use of astrometric as well as new and
existing spectroscopic observations, to present improved dynamical mass
estimates for another binary system in the Hyades, \vstar. This is the
ninth such example in the cluster. With formal mass uncertainties for
the primary and secondary under 1.8\%, it ranks among the best
determinations in this small group. We also derived the orbital
parallax of the system to better than 0.7\%, and presented evidence of
its superior accuracy (and precision) compared to Gaia~DR3, based on
independent kinematic parallax determinations by others.

These observational results show relatively good agreement in the mass-luminosity
diagram with two sets of current stellar evolution models (MIST, and
PARSEC~v1.2S) for the known age and metallicity of the cluster,
in both the $V$ and $
K$ bandpasses. Only one other Hyades system is
available for this type of comparison in the $K$ band
\citep[HD~284163;][]{Torres:2023}.

Even though \vstar\ is not eclipsing, we have also obtained estimates
of the absolute radii and effective temperatures of the
components through a spectral energy distribution analysis,
supplemented by the orbital parallax. Those results are also
consistent with theory, with the temperatures slightly favoring the
MIST models.

Examination of the TESS photometry for \vstar\ shows variability at a
low level ($\leq 1.5$~mmag total amplitude), with a period of about
8.5~d that is similar to a previous estimate from the literature, and
is presumably due to rotation combined with variable surface activity
on one or both stars.

\begin{acknowledgements}

The spectroscopic observations of \vstar\ at the CfA were obtained
with the help of J.\ Caruso and J.\ Zajac. We thank
R.\ Davis for maintaining the Digital Speedometer
database over the years, and M.\ McEachern (Wolbach Library)
for assistance with publications that are not accessible online.
{We also thank the anonymous referee for helpful comments.}

This research has benefited from the use of the SIMBAD and VizieR
databases, operated at the CDS, Strasbourg, France, and of NASA's
Astrophysics Data System Abstract Service. The computational resources
used for this research include the Smithsonian High Performance
Cluster (SI/HPC), Smithsonian Institution
(\url{https://doi.org/10.25572/SIHPC}).

\end{acknowledgements}

\end{document}